# The impact of accumulative pension policy on welfare of individuals

**Marika Khozrevanidze** [1]

Shota Rustaveli State University, Batumi, Georgia

**Abstract**

Many countries around the world have had to carry out radical reforms periodically in their pension systems.Global experience shows that it is important to optimize the costs of pension and social security systems in order to ensure a decent old age in addition to reducing the pressure on budgetary resources. By Georgia's changing demographic situation, special attention is paid to proper functioning of the pension policy. The pension reform carried out in Georgia in 2019 caused a difference of opinion among experts. This issue in today's conditions does not lose relevance. The presented thesis discusses the impact of the mandatory funded pension system on the well-being of people. Thesis includes the following issues: peculiarities of the formation of pension systems in Georgia. It is presented a small historical excursion in terms of the development of pension systems. In addition, are discussed the international experience of pension systems and comparative analysis in relation to Georgia. This paper specifically focuses on the essence of the mandatory funded pension system and assesses the current situation in terms of investment potential of the resource accumulated in the pension fund. In conclusion, are presented the challenges of this system and the ways of perfection.

---

[1] Scientific supervisor: G. Abuselidze, Doctor of Economics, Professor

# დაგროვებითი საპენსიო პოლიტიკის გავლენა ფიზიკურ პირთა კეთილდღეობაზე

მარიკა ხოზრევანიძე[2]
ბათუმის შოთა რუსთაველის სახელმწიფო უნივერსიტეტი, საქართველო

## ანოტაცია

საპენსიო სისტემებში რადიკალური რეფორმების გატარება პერიოდულად მსოფლიო მასშტაბით უამრავ ქვეყანას უწევდა. გლობალური გამოცდილება ცხადყოფს, რომ მნიშვნელოვანია საპენსიო და სოციალური დაცვის სისტემების ხარჯების ოპტიმიზაცია, რათა საბიუჯეტო რესურსებზე არსებული წნეხის შემცირების პარალელურად უზრუნველყოფილი იქნეს ადამიანების ღირსეული სიბერე. საქართველოში არსებული ცვალებადი დემოგრაფიული მდგომარეობის გათვალისწინებით, განსაკუთრებული ყურადღება ექცევა გამართული საპენსიო პოლიტიკის ფუნქციონირებას. 2019 წელს საქართველოში გატარებულმა საპენსიო რეფორმამ ექსპერტთა შორის აზრთა სხვადასხვაობა გამოიწვია. აღნიშნული საკითხი დღევანდელ პირობებშიც არ კარგავს აქტუალურობას. წარმოდგენილი ნაშრომი კვლევითი ხასიათისაა და განხილულია სავალდებულო დაგროვებითი საპენსიო სისტემის გავლენა ფიზიკურ პირთა კეთილდღეობაზე. ნაშრომში წარმოდგენილია საპენსიო სისტემების ფორმირების თავისებურებები საქართველოში და მცირე ისტორიული ექსკურსი საპენსიო სისტემების განვითარების თვალსაზრისით. ამასთან, განხილულია საპენსიო სისტემების საერთაშორისო გამოცდილება და შედარებითი ანალიზი საქართველოსთან მიმართებაში. ნაშრომში მნიშვნელოვანი ყურადღება გამახვილებულია სავალდებულო დაგროვებითი საპენსიო სისტემის არსზე და შეფასებულია არსებული მდგომარეობა საპენსიო ფონდში აკუმულირებული რესურსის საინვესტიციო პოტენციალის თვალსაზრისით. დასკვნის სახით კი წარმოდგენილია აღნიშნული სისტემის გამოწვევები და სრულყოფის გზები.

საკვანძო სიტყვები: საპენსიო პოლიტიკა, კეთილდღეობა, ბიუჯეტი, საქართველო

## შესავალი

ქვეყნის სოციალური პოლიტიკის ერთ-ერთი მნიშვნელოვანი განმსაზღვრელი ინდიკატორი საპენსიო სისტემაა. მისი გამართული ფუნქციონირების გარეშე წარმოუდგენელიც კია ქვეყნის სრულფასოვანი ეკონომიკური მდგრადობა.

განვითარებადმა ქვეყნებმა, მათ შორის საქართველომ, გააცნობიერა, რომ აუცილებელია საპენსიო სისტემის რეფორმირება. საპენსიო სისტემამ უნდა უპასუხოს ქვეყნის წინაშე მდგარ, როგორც ფინანსურ, ასევე ეკონომიკურ და დემოგრაფიულ გამოწვევებსაც. სიცოცხლის ხანგრძლივობის ზრდის პარალელურად გახშირებულმა ფინანსურმა კრიზისებმა ჯერ კიდევ 90-იანი წლებიდან დღის წესრიგში დააყენა საპენსიო სისტემების სრულყოფის საკითხი.

მსოფლიო ბანკის ანგარიშიდან ირკვევა, რომ განვითარებულ და ძირითადად განვითარებად ქვეყნებშიც სულ უფრო და უფრო იზრდება ხანდაზმული მოსახლეობის წილი. ეს უკანასკნელი კი გულისხმობს ქვეყნებში საპენსიო უზრუნველყოფის ხარჯების მნიშვნელოვან ზრდას-დამატებით წნეხს ერთი მხრივ, საჯარო ფინანსებზე ანუ საბიუჯეტო სახსრებზე და მეორე მხრივ, გადასახადის გადამხდელ მოსახლეობაზე.

---

[2] მეცნიერ-ხელმძღვანელი: გ. აბუსელიძე, ეკონომიკის დოქტორი, პროფესორი

საქართველოში 2019 წლის 1 იანვრიდან ამოქმედდა სავალდებულო დაგროვებითი საპენსიო სისტემა, რომელსაც მნიშვნელოვანი როლი უნდა შეესრულებინა ქვეყნის ეკონომიკურ-სოციალურ პოლიტიკაში. რეფორმის მთავარი მიზანი იყო თანამედროვე და ლიბერალური ეკონომიკური გარემოს შექმნა, რომელსაც ექნებოდა ადეკვატური რეაქცია გახშირებულ ფინანსურ კრიზისებზე.

აქედან გამომდინარე, შეგვიძლია დავასკვნათ რომ საქართველოში აღნიშნული რეფორმის გატარება გამოუვალი მდგომარეობით დამდგარი აუცილებლობა იყო.

ნაშრომის კვლევის მიზანია ფიზიკურ პირებზე საპენსიო რეფორმის შედეგების განსაზღვრა, ასევე საპენსიო ფონდში აკუმულირებული რესურსის ფორმირების პროცესი და მისი გამოყენების პერსპექტივების ანალიზი. აღნიშნული მიზნებიდან გამომდინარე კვლევის ამოცანაა საერთაშორისო გამოცდილების განხილვა, არსებული საპენსიო მოდელების SWOT ანალიზი, საქართველოში არსებული საპენსიო სისტემის ნაკლოვანებებისა და შესაძლებლობების წარმოჩენა. გარდა ამისა, ნაშრომის ამოცანაა მოძიებულ ინფორმაციაზე და რესურსებზე დაყრდნობით მოვახდინო არსებული საპენსიო მოდელის სრულყოფის მექანიზმების განხილვა და რეკომენდაციების შემუშავება.

ნაშრომის ეფუძნება საპენსიო სისტემების მოდელებზე, საპენსიო სისტემის სრულყოფის მექანიზმებზე და საპენსიო სისტემაში განხორციელებულ რეფორმებზე შექმნილ ქართველი, თუ უცხოელი ეკონომისტებისა და ფინანსისტების ნაშრომებს. გამოყენებულია თემატიკასთან დაკავშირებით სამთავრობო თუ არასამთავრობო სექტორის მიერ გამოქვეყნებული სტატიები და ანგარიშები. ნაშრომში ასევე გამოყენებული საპენსიო სააგენტოს მიერ მოწოდებული საჯარო ინფორმაცია აკუმულირებულ რესურსებთან დაკავშირებით. გარდა ამისა, კვლევისათვის გამოყენებულ იქნა თვისობრივი და რაოდენობრივი კვლევის მეთოდები: დოკუმენტების ანალიზი, ექსპერტთა აზრის შესწავლა სტატიებზე და ინტერვიუებზე დაყრდნობით და საზოგადოებრივი აზრის კვლევა დახურული ონლაინ ფორმატის კითხვარით.

1. საპენსიო სისტემების მოდელების თავისებურებები
*1.1 საპენსიო სისტემების ფორმირების თავისებურებები საქართველოში*

ოდითგანვე, საპენსიო სისტემების სრულყოფის მთავარი მიზანი ხანდაზმული ადამიანებისთვის სიღარიბის შემცირება და მათთვის ღირსეული პირობების შექმნა იყო. დღესდღეისობით, ნებისმიერი ქვეყნის სოციალურ-ეკონომიკური მდგრადობის შეფასების ერთ-ერთ მთავარ ინდიკატორად შეგვიძლია მივიჩნიოთ საპენსიო პოლიტიკა.

ჯერ კიდევ მე-20 საუკუნიდან ევროპის ქვეყნებში შეიმჩნეოდა დემოგრაფიული გამოწვევები, რაც გულისხმობს მოსახლეობის სიცოცხლის საშუალო ხანგრძლივობის ზრდასა და შობადობის შემცირებას. აღნიშნული ცნობილია ტერმინით - მოსახლეობის დაბერება. ეს უკანასკნელი იწვევდა სოციალური ხარჯების ზრდას, რომელიც მეტად ართულებდა შრომის ასაკში მყოფი მოსახლეობის მდგომარეობას. ბუნებრივი მატების კოეფიციენტი საქართველოში 2014 წლის შემდეგ კლების ტენდენციით ხასიათდება, უკვე 2020 წლის მონაცემებით უარყოფითია და შეადგენს -1,1%-ს. (იხ. დიაგრამა #1)

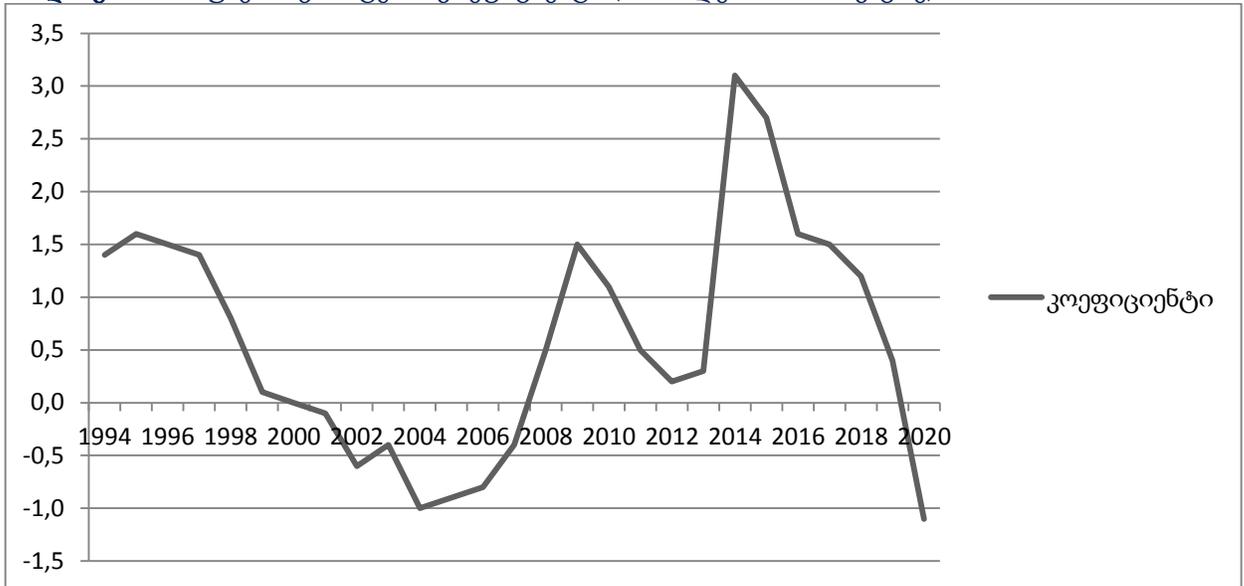

დიაგრამა 1: ბუნებრივი მატების კოეფიციენტი (მოსახლეობის 1000 კაცზე)

წყარო: საქართველოს სტატისტიკის ეროვნული სამსახური ეს გახლდათ მთავარი ბიძგი საპენსიო სისტემაში მრავალი რეფორმის გატარების, გამონაკლისი არც საქართველო აღმოჩნდა, რომელსაც საპენსიო პოლიტიკაში არაერთი ცვლილების შეტანა მოუწია სხვადასხვა ეკონომიკური თუ პოლიტიკური ფაქტორებიდან გამომდინარე.

საქართველოში არსებული საპენსიო სისტემის მიმოხილვისთვის ავლის წერტილად ავიღოთ დამოუკიდებლობის მოპოვების თარიღი, როდესაც საქართველოში საბჭოთა საპენსიო სისტემა ფუნქციონირებდა. მისი უმთავრესი მიზანი და ამოცანა პენსიაზე გასული ადამიანისთვის მანამდე არსებული ცხოვრების დონის შენარჩუნება გახლდათ. მსოფლიო ბანკის მონაცემებით, 1993 წელს აღნიშნული სისტემის ფარგლებში პენსია ბოლო სამუშაო წლის (ან ბოლო 5 წლის) ხელფასის 55%-ს შეადგენდა.

შექმნილი იყო ერთიანი საპენსიო ფონდი, რომელიც ფინანსდებოდა სოციალური შენატანებისაგან, კერძოდ, დამსაქმებელი იხდიდა სახელფასო ფონდის 37%-ს, საბიუჯეტო ორგანიზაციები ასევე სახელფასო ფონდის 26%-ს ხოლო უშუალოდ დასაქმებული ხელფასის 1%-ს. (The World Bank, 1993). აღნიშნული ფონდი ახდენდა პენსიებისა და სხვა ყველა სოციალური გასაცემლის ადმინისტრირებას.

დამოუკიდებლობის მოპოვების შემდგომ ქვეყანაში რთული ეკონომიკური და პოლიტიკური სიტუაცია შეიქმნა. ეკონომიკურმა კრიზისმა და არასტაბილურმა ფისკალურმა გარემომ, რა თქმა უნდა, გავლენა იქონია საპენსიო ვალდებულებებზე, რომლის მოცულობაც ამ პერიოდისთვის შეადგენდა მთლიანი შიდა პროდუქტის 10%-ს.

საპენსიო დანახარჯების ეს რაოდენობა საგანგაშო აღმოჩნდა, რამდენადაც მსოფლიო ბანკის მიერ გამოქვეყნებული 1993 წლის ანგარიშის მიხედვით, საპენსიო ფონდში დაგეგმილი შემოსავლების მხოლოდ 50%-60% აკუმულირდა. პენსიების გაცემა არასრულად და არარეგულარულად ხორციელდებოდა.

სოციალური გადასახადის, იგივე შენატანების, განაკვეთი ეტაპობრივად იცვლებოდა. 2004 წლისთვის ქვეყანაში ლიბერალური ეკონომიკური პოლიტიკა გატარდა, ეკონომიკური აქტივობების წახალისების მიზნით კი სოციალური შენატანების განაკვეთი შემცირდა 20%-მდე. 2008 წელს სოციალური გადასახდი ამ ფორმით გაუქმდა, სანაცვლოდ კი ცვლილება შევიდა საშემოსავლო გადასახადის საპროცენტო განაკვეთში და ის 12%-დან 20%-მდე გაიზარდა. აღნიშნული ცვლილების შემდეგ ფიზიკური პირებისთვის პენსიის მოცულობა აღარ იყო დამოკიდებული პენსიაზე გასვლამდე არსებულ შემოსავლებზე. პენსიის მოცულობის განსაზღვრა სრულიად სახელმწიფომ აიღო საკუთარ თავზე.

ასევე, 2004 წლიდან დაიწყო პენსიის ზრდის ტენდენცია, მანამდე 1998 წლიდან მოყოლებული ის შეადგენდა 14 ლარს. ეს თანხა მნიშვნელოვნად ჩამოუვარდებოდა საარსებო მინიმუმს და საშუალო ხელფასის 11,1%-ს შეადგენდა. აქვე უნდა აღინიშნოს ისიც რომ, წლების განმავლობაში ინფლაციის გამო აღნიშნული 14 ლარი სულ უფრო და უფრო კარგავდა მსყიდველობითუნარიანობას.

2007 წლიდან საქართველოს საპენსიო სისტემაში შემდეგი ეტაპი იწყება. ეს უკანასკნელი მოიცავდა სახელმწიფო პენსიის დანამატს სტაჟის მიხედვით:

ცხრილი #1 პენსიის დანამატი შტაჟის მიხედვით

| საერთო შრომითი სტაჟი | საპენსიო დანამატი |
|---|---|
| 5 წლამდე | 2 ₾ |
| 5-15 წლამდე | 4 ₾ |
| 15-25 წლამდე | 7 ₾ |
| 25 და მეტი | 10 ₾ |

წყარო: საქართველოს მთავრობის დადგენილება №181[3]

ეს წესი გაუქმდა 2012 წელს, მას შემდეგ, რაც ყველა პენსიონერისთვის გათანაბრდა პენსია და გახდა 110 ლარი 67 წლამდე ასაკი ადამიანებისთვის და 125 ლარი 67 წელს ზემოთ მყოფი პირებისთვის.

საპენსიო სისტემის ფორმირების პროცესში მესამე მნიშვნელოვანი და გარდამტეხი ეტაპი დაიწყო 2014 წლიდან, როდესაც სოციალური უზრუნველყოფის ერთიანმა სახელმწიფო ფონდმა დაკარგა თავისი ძირითადი ფუნქცია: სოციალური გადასახადის

---
[3] საქართველოს მთავრობის დადგენილება №181- შრომითი სტაჟის მიხედვით საპენსიო ასაკის საფუძვლით დანიშნული სახელმწიფო პენსიის დანამატის განსაზღვრის შესახებ. shorturl.at/klnBR

შეგროვება და ადმინისტრირება, შედეგად, განხორციელდა მისი სრული რეორგანიზაცია და შეიქმნა ორი ახალი სტრუქტურა:

- დასაქმებისა და სოციალური უზრუნველყოფის სახელმწიფო სააგენტო
- ჯანდაცვისა და სოციალური პროგრამების სააგენტო

2010 წელს აღნიშნული სააგენტოები გაერთიანდა და ჩამოყალიბდა სოციალური მომსახურების სააგენტო, რომელსაც დაეკისრა სახელმწიფო პენსიის გამოყოფის პასუხისმგებლობა, იგი შრომის, ჯანდაცვისა და სოციალური დაცვის სამინისტროს ერთეულს წარმოადგენდა. შეგვიძლია ვთქვათ, რომ აღნიშნული რეფორმა სოციალური პროგრამების (რომელიც მოიცავდა პენსიებსაც) დაფინანსების კუთხით განხორციელებული ერთ-ერთი ყველაზე მნიშვნელოვანი აღმოჩნდა.

სოციალური პენსიის რაოდენობა არის მნიშვნელოვანი ინსტრუმენტი სოციალური დაცვის კუთხით. მაგრამ, იგი სრულად ვერ განაპირობებს მაღალი სტანდარტის ცხოვრების პირობებს. აქედან გამომდინარე, აუცილებელია, დამატებითი შემოსავლის მიღების მიზნით, კერძოდ, დანაზოგების სისტემის დანერგვა, რადგან პენსიაზე გასვლის შემდეგ მოქალაქეებს გაუჩნდეთ დამატებითი შემოსავალი, უფრო მაღალი ცხოვრების სტანდარტის უზრუნველსაყოფად.

როდესაც ვსაუბრობთ საპენსიო ასაკს იდწეული ფიზიკური პირებისათვის კეთილდღეობის უზრუნველყოფაზე აუცილებელია აღვნიშნოთ ე.წ. ჩანაცვლების კოეფიციენტის[4] მნიშვნელობა. კოეფიციენტი გვიჩვენებს პენსიისა და ხელფასის ფარდობას. ქვემოთ მოცემული ცხრილი კი გვიჩვენებს კოეფიციენტის მიხედვით საქართველოსა და სხვა ქვეყნებს შორის სხვაობას.

ცხრილი #2 ქვეყნების კლასიფიკაცია ჩანაცვლების კოეფიციენტის მიხედვით

| ქვეყანა | ჩანაცვლების კოეფიციენტი |
|---|---|
| საქართველო | 18% |
| პოლონეთი | 29,4% |
| გერმანია | 38.7% |
| სლოვაკეთი | 38,8% |
| OECD[5] საშუალო | 49% |
| ევროკავშირის ქვეყნები | 52% |
| წყარო: Net pension replacement rates, oecd-ilibrary[6] | |

2021 წლის 1 იანვრის მდგომარეობით საქართველოს 3 728,6 ათას კაცს შეადგენს. აქედან ასაკით პენსიის მიმღები არის 783 705 კაცი. (იხ. დიაგრამა #2)

---

[4] პენსიის ჩანაცვლების კოეფიციენტი - პენსიის ოდენობის შედარება ხელფასთან, რომელიც პირს ჰქონდა პენსიაზე გასვლამდე.

[5] OECD ეკონომიკური თანამშრომლობისა და განვითარების ორგანიზაცია. ინგლ. Organization for Economic Co-operation and Development
განვითარებული ქვეყნების, რომლებიც ცნობენ წამომადგენლობითი დემოკრატიისა და თავისუფალი საბაზრო ეკონომიკის პრინციპებს. დღესდღეობით მისი წევრია 37 ქვეყანა.

[6] Net pension replacement rates -shorturl.at/ituNZ

**დიაგრამა #2:** ასაკით პენსიის მიმღებთა რაოდენობა 2007-2020 წლებში

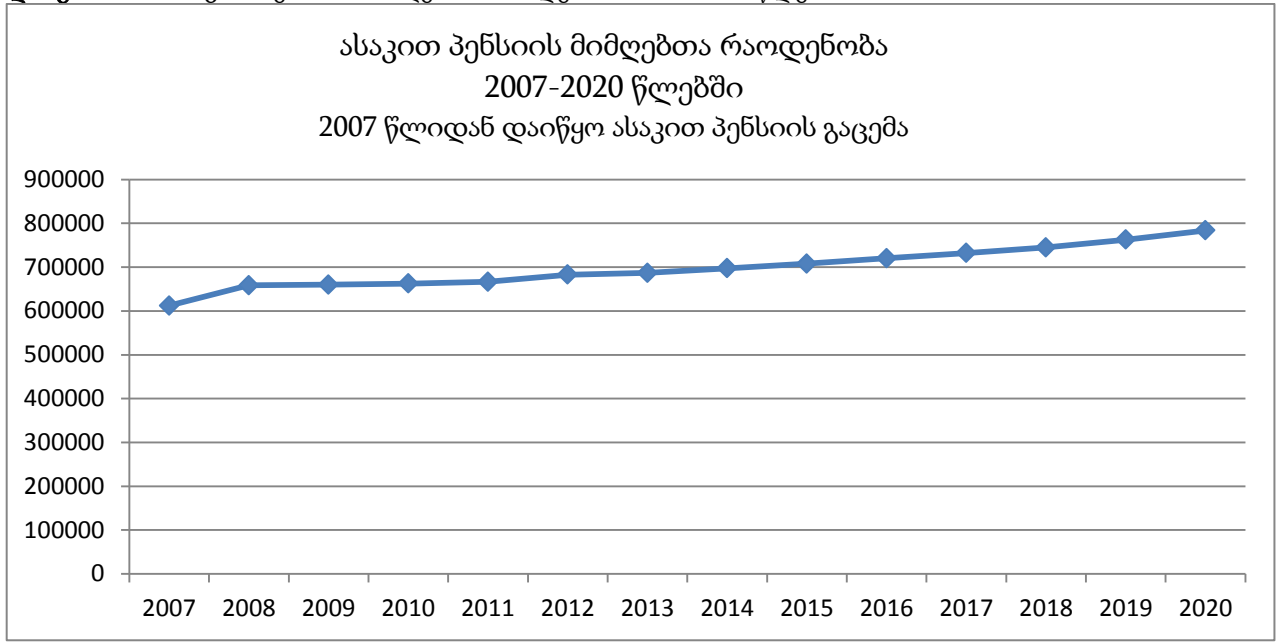

წყარო: საქართველოს სტატისტიკის ეროვნული სამსახური

„2015 წლის შემდგომ ნათლად იკვეთება მნიშვნელოვანი დემოგრაფიული პრობლემა. ყოველწლიურად იზრდება ასაკით პენსიონერთა რიცხვი, საქართველოს მოსახლეობის ასაკობრივი სტრუქტურა შეიძლება ითქვას, რომ დაბერებას განიცდის. დემოგრაფიულ ანალიზზე დაყრდნობით, ვარაუდობენ, რომ 2025 წლის შემდგომ პენსიონერთა რაოდენობა გადააჭარბებს ბავშვების რაოდენობას. ამის პარალელურად კი შემცირდება შრომისუნარიანი მოსახლეობის, მათ შორის გადასახადის გადამხდელი პირების რაოდენობის ზრდა." [საქართველოს ეკონომიკისა და მდგრადი განვითარების სამინისტრო; საქართველოს საპენსიო რეფორმა][7] ამავე სტატიიდან ირკვევა, რომ 2030 წელს პენსიის მიმღებთა რაოდენობა ერთ მილიონს მიაღწევს, რაც იმას ნიშნავს, რომ ასაკოვანი მოსახლეობის რაოდენობა სრული მოსახლეობის მეოთხედზე მეტი იქნება. დროის ასე მცირე მონაკვეთში მსგავსი მონაცემები უმნიშვნელოვანეს დემოგრაფიულ ცვლილებას წარმოადგენს.

აღნიშნული მზვავი პრობლემის შედეგი შესაძლებელია აღმოჩნდეს ხანდაზმულთა შორის უკვე არსებული სიღარიბის კიდევ უფრო გაღრმავება. პენსიის რაოდენობა საქართველოში სხვა ცივილიზებულ ქვეყნებში არსებულ პენსიას საგრძნობლად ჩამორჩება. შესაბამისად, იგი ვერ უზრუნველყოფს არა თუ მაღალი, არამედ საშუალო სტანდარტის ცხოვრების პირობებსაც კი.

---
[7] საქართველოს ეკონომიკისა და მდგრადი განვითარების სამინისტრო; საქართველოს საპენსიო რეფორმა
shorturl.at/vyQR8

2019 წლამდე საქართველოში არსებული საპენსიო სისტემა ფინანსდებოდა ერთიანი სახელმწიფო ბიუჯეტიდან, რომლის უმნიშვნელოვანესი წილი წარმოდგენილია დასაქმებული ადამიანების მიერ გადახდილი საშემოსავლო გადასახადით. პენსიის შესახებ საქართველოს კანონის მიხედვით, საქართველოს ყველა მოქალაქეს საპენსიო ასაკის მიღწევისთანავე (მამაკაცებისთვის 65 წელი და ქალებისათვის 60 წელი) უფლება აქვს, მოთხოვოს სოციალური პენსია. გავრცელებული პრაქტიკების მიხედვით, იდენტური მახასიათებლების მქონდე საპენსიო სისტემები ცნობილია, როგორც ნულოვანი პილარი, რომლის მიზანიცაა სრული მოსახლეობის თანაბარი ოდენობით პენსიით უზრუნველყოფა. საერთაშორისო სტანდარტის მიხედვით, ნულოვანი პილარი, იგივე სოციალური პენსია გვხვდებოდა შემდეგ ქვეყნებში: კანადა, ახალი ზელანდია, ჩილე, ყაზახეთი და ა.შ.

ადამიანები თანხმდებიან, რომ პენსია წარმოადგენს სიღარიბესთან ბრძოლის უმნიშვნელოვანეს ბერკეტსა თუ მექანიზმს. საპენსიო სისტემაში რეფორმების გატარების აუცილებლობაზე მსჯელობა კი საქართველოში დიდი ხანია დაწყებულია.

*1.2 საპენსიო სისტემების მსოფლიო გამოცდილება და შედარებითი ანალიზი*

საპენსიო სისტემების ფორმირებაში ძირეული ძვრები მე-19 საუკუნის მეორე ნახევრიდან დაიწყო. მანამდე სოციალურად შეჭირვებული ადამიანების და ჯგუფების დამხმარის როლში ძირითადად წარმოდგენილი იყო ოჯახები, სხვადასხვა რელიგიური გაერთიანებები და დაწესებულებები, თემები და ა.შ. რამდენადაც, აგრარული საზოგადოების წარმომადგენლებისათვის დამახასიათებელი იყო დიდი ოჯახები (რამდენიმე თაობისაგან შემდგარი) მათვის ჩვეულებრივი მოვლენა გახლდათ ოჯახის უფროსი წარმომადგენლებისათვის დახმარების გაწევა.

მე-19 საუკუნის მეორე ნახევრიდან სოციალური რისკების მართვის სადავეები სახელმწიფომ აიღო საკუთარ ხელში. ამ პერიოდიდან ევროპაში უკვე გვხვდება არაერთი სოციალური დაცვის სისტემა, რომლებიც სხვადასხვა რისკებისგან აზღვევდნენ მოსახლეობას, მათ შორის ხანდაზმულობა, დაავადებები, სიღარიბე, შეზღუდული შესაძლებლობები და ა.შ.

სწორედ აღნიშნული პერიოდიდან გვხვდება მიგრაციული პროცესების დაწყებაც, კერმოდ, სამუშაო ძალის სოფლებიდან ქალაქებისკენ მიგრაცია. ადამიანები იძულებულნი გახდნენ ემუშავათ ნათესაური თუ სოციალური კავშირების გარეშე. ამ ყოველივემ კი გამოიწვია სიბერეში შესაბამისი ფინანსური უზრუნველყოფის გარეშე დარჩენის რისკის ზრდა. საპენსიო სისტემების დანერგვა სწორედ ამ პერიოდიდან იღებს სათავეს.

1889 წელს გერმანიის კანცლერი ოტო ფონ ბისმარკი გამოვიდა ინიციატივით გერმანიაში სიბერისა და შრომისუნარიანობის დაზღვევის კანონის მიღებასთან დაკავშირებით. სწორედ გერმანია იყო პირველი ქვეყანა სადაც სოციალური დაზღვევის სისტემა პირველად შეიქმნა. აღნიშნული წესი ეხებოდა მხოლოდ იმ მუშებს, რომელთა შემოსავალი არ აღემატებოდა წინასწარ დადგენილ ზღვრებს. სოციალური დაზღვევის სისტემას ფინანსურად უზრუნველყოფდა როგორც სახელმწიფო, ისე დასაქმებული და დამსაქმებელი. მთავარი განსხვავება იყო ის, რომ სახელმწიფოს შენატანი იყო ფიქსირებული სისტემის ყველა მონაწილისათვის, მიუხედავად მათი შემოსავლებისა. ხოლო რაც შეეხება დასაქმებულსა და დამსაქმებელს, ისინი ხელფასის პროპორციულად აკეთებდნენ შენატანებს.

იმისათვის რომ პირს მიეღო პენსია, აუცილებელი იყო

- 70 წელს (საპენსიო ასაკი) მიღწევა
- შრომის უნარიანობის დაკარგვა

აღნიშნული სისტემის მთავარი მიზანი იყო, მოსახლეობისთვის პენსიაზე გასვლის შემდგომ შეენარჩუნებინათ ცხოვრების ის პირობები, რაც მათ პენსიაზე გასვლამდე ჰქონდათ.

მეორე მსოფლიო ომამდელ პერიოდში ბისმარკის აღნიშნული საპენსიო სისტემის მოდელი გავრცელდა ევროპის სხვა ქვეყნებშიც, მათ შორის: შვედეთში, ნიდერლანდებში, ლუქსემბურგში და ა.შ. თუმცა, ასევე უნდა აღინიშნოს, რომ ბისმარკისეული მოდელის მოდერნიზაცია სხვადასხვა ქვეყანაში განსხვავებული სცენარით წარიმართა. მაგალითად ლათინო ამერიკის რიგ ქვეყნებში პენსიის მიმღებთა წრე გარკვეულწილად შეზღუდული იყო საქმიანობის სფეროების მიხედვით: არგენტინაში პენსიას მხოლოდ საჯარო მოხელეები იღებდნენ, ან ბანკში დასაქმებული პირები. ასევე გამოყოფილი იყო რკინიგზაზე და პორტში დასაქმებული პირებიც. განსხვავებული პრაქტიკა გვხვდება მხოლოდ ჩილეში, სადაც საპენსიო სისტემაში ჩართული პირების საქმიანობა არ იყო შემოსაზღვრული არანაირი ჩარჩოებით. ნებისმიერი ადამიანი, თუ მისი შემოსავალი დადგენილ ზღვარზე დაბალი იყო ავტომატურად ერთვებოდა სისტემაში.

ბისმარკის მოდელისგან სრულიად განსხვავებული აღმოჩნდა ე.წ შენატანებზე დაფუძნებული სოციალური დაზღვევის სისტემა, რომელიც პირველად დიდ ბრიტანეთსა და დანიაში დაინერგა. აღნიშნული სისტემის მიზანი არა პენსიონერებისთვის არსებული დონის შენარჩუნება, არამედ სიღარიბის დონის შემცირება.

1990 წლის მეორე ნახევრიდან გახშირდა მსხვილი მასშტაბების ეკონომიკური და ფინანსური კრიზისები, რამაც ბევრ ქვეყანაში მნიშვნელოვანი პრობლემები გაუჩინა საჯარო ფინანსებსა თუ კერძო საპენსიო ფონდებს. ამას დაემატა მსოფლიო საპენსიო სისტემაში ხანდაზმულების ახალი და მოცულობითი ტალღა, რომელთა ფინანსურად უზრუნველყოფა მნიშვნელოვან რესურსებს საჭიროებდა.

მოსახლეობის ზრდის ძალიან მაღალი ტემპების პარალელურად იზრდებოდა სამუშაო ძალის რაოდენობაც. ყოველივე ამან გამოიწვია დიდი მოცულობის საპენსიო რეზერვების შეგროვება. სახელმწიფოებს გაუჩნდათ მოტივაცია გადასულიყვნენ გადანაწილებით საპენსიო სისტემაზე (pay as you go). ამ სისტემის მიხედვით, საპენსიო შენატანების ავტორები აღარ აგროვებდნენ თანხებს სამომავლო მოხმარებისათვის, ეს რესურსი მიემართებოდან ხანდაზმულების საპენსიო უზრუნველყოფისაკენ. სანაცვლოდ, საპენსიო შენატანების ავტორებს ჰქონდათ გარანტია, რომ სახელმწიფო მათ გადაუხდიდა გარკვეული რაოდენობის პენსიას სამოვალოდ. ეს პერიოდი კიდევ იმით იყო გამორჩეული, რომ პენსიის რაოდენობა იმ ნიშნულამდე გაიზარდა, რომ ადამიანებს აღარ უწევდათ პენსიის პარალელურად მუშაობა თავის რჩენისათვის, რადგან საპენსიო შენატანებით მათ დაინსახურეს ღირსეული, შრომისგან თავისუფალი ცხოვრება.

თანამედროვე სამყაროში ყველაზე მეტად გავრცელებულია დასავლეთ ევროპული საპენსიო სისტემა, რომელიც ძირითადად ეყრდნობა ე.წ. სვეტებიან სისტემას. ამგვარი საპენსიო სქემები განსხვავდებიან სისტემაში მონაწილე პირების რაოდენობით, მათი დაცულობის ინდექსით და დაფინანსების პრინციპით.

- პირველი სვეტი - მოიცავს სოლიდარობის პრინციპზე დაფუძნებულ საპენსიო მოდელს, რომლითაც სხვადასხვა სოციალური ჯგუფების საპენსიო უზრუნველყოფა ხდება. ხშირ შემთხვევაში ეს არის ფიქსირებული, გარანტირებული თუმცა მინიმალური თანხა.
- მეორე სვეტი - აღნიშნული სვეტი წარმოადგენს ან სავალდებულო ან ნებაყოფლობით საპენსიო სქემებს, რომლის დროსაც ერთიანდება საპენსიო შენატანები ვალდებულებებზე დანამატის სახით. ამის მაგალითია დაგროვებითი საპენსიო სისტემა.
- მესამე სვეტი - მისი მთავარი დამახასიათებელია ნებაყოფლობითობის პრინციპი. ერთიანდება ინდივიდუალური შენატანები სხვადასხვა კერძო საფინანსო ინსტიტუტებში.

განვიხილოთ პოლონეთის შემთხვევა.

პოლონეთის მაგალითის განხილვა და საქართველოსთან შედარებითი ანალიზი მიზანშეწონილია ორი მთავარი მიზეზის გამო: ორივე ყოფილი სოციალისტური ბანაკის წევრი ქვეყანაა და მეორე პოლონეთი იყო ერთ-ერთი პირველი ქვეყანა, რომელმაც სავალდებულო დაგროვებითი საპენსიო სისტემა დანერგა.

პოლონეთი იყო ერთ-ერთი პირველი ცენტრალური და აღმოსავლეთ ევროპის ქვეყანა, რომელმაც საკუთარი საპენსიო სისტემის სრული პრივატიზაცია განახორციელა. მკვლევართა აზრით, ამ ყოველივემ სხვადასხვა ეკონომიკურ ფაქტორებთან ერთად არაერთი სოციალური შედეგი მოჰყვა, როგორიცაა: განახლებული მენტალიტეტი და ახალი შრომის დისციპლინა, მისი მთავარი მიზანი თვითრეგულირებადი ადამიანის შექმნა იყო, რომელიც მარტივად და დანაკარგების გარეშე ადაპტირდებოდა საბაზრო ცვლილებებთან.

1999 წელს პოლონეთში დაიწყო საპენსიო რეფორმების ციკლი, მანამდე მოქმედებდა მხოლოდ სახელმწიფო განაწილებითი საპენსიო სისტემა (pas as you go). 2002 წლიდან პოლონეთში შეიქმნა დემოგრაფიული სარეზერვო ფონდი, რომლის მთავარი მიზანი მომავალში შესაძლო დეფიციტის რისკისგან დაზღვევა. დღესდღეისობით, პოლონეთის საპენსიო სისტემა მიჩნეულია, როგორც ერთ-ერთი ყველაზე მდგრადი საპენსიო მოდელი მთელი ევროპის მასშტაბით.
პოლონეთში მოქმედებს პენსიის სამსვეტიანი სისტემა.

I დონე, სახელმწიფო პენსია, წარმოადგენს სავალდებულო სოლიდარულ სქემას, რომელიც დამყარებულია განსაზღვრულ შენატანებზე. მკვაცრადაა განსაზღვრული პენსიონერებისათვის გასაცემი თანხები შენატანების მიხედვით. სისტემის ყველა მონაწილეს ესნება ანგარიში, სადაც ასახულია მთელი ცხოვრების განმავლობაში გაკეთებული ფინანსური კონსტრიბუცია. საპენსიო სარგებლის გამოთვლის დროს ყურადღება ექცევა იმ პერიოდისათვის სიცოცხლის საშუალო ხანგრძლივობას. პარალელურად სახელმწიფო უზრუნველყოფს პენსიაზე ნაადრევად გასვლის ხელშემწყობი ფაქტორების მაქსიმალურად გამორიცხვას. საპენსიო ფონდში შესატანი თანხა თანაბრად ნაწილდება დასაქმებულსა და დამსაქმებელზე. თუ აღნიშნული სქემის მიხედვით პენსიონერის პენსია საშუალოზე დაბალი იქნება, მაშინ სახელმწიფო სოლიდარული ფონდიდად უზრუნველყოფს მას მინიმალური პენსიით.

II დონე მოიცავს სავალდებულო ინდივიდუალურ საპენსიო ანგარიშებს. მოცემული დონე მსგავსია ღია საპენსიო ფონდების სტრუქტურის. საპენსიო საფონდო სისტემებს აქვთ საქციო საზოგადოების იურიდიული ფორმა. რა თქმა უნდა, მსგავსი ფოდნების შესაქმნელად საჭიროა სპეციალური ნებართვა სადაზღვეო და საპენსიო ფონდების ზედამხედველობის კომისიისაგან (KNF)[8]. მისი სტრუქტურის მთავარი შემადგენელი ნაწილია საზედამხედველო საბჭო და საერთო კრება.

აკუმულირებული რესურსების სხვადასხვა საინვესტიციო მიმართულებით განკარგვა შეზღუდულია შემდეგნაირად:

ცხრილი #3: საპენსიო ფონდში აკუმულირებული რესურსების განკარგვის მიმართულებები პოლონეთში

| არაუმეტეს 40% | საფონდო ბირჟა; |
| არაუმეტეს 40% | იპოთეკური, მუნიციპალური, კორპორატიული ობლიგაციები; |
| არაუმეტეს 10% | რეგულირებადი, არასავალუტო ბაზარი; |
| არაუმეტეს 10% | საბანკო დეპოზიტები |

წყარო: pensionfundsonline[9]

---

[8] KNF- Polish: Komisja Nadzoru Finansowego; Financial Supervision Commission-პოლნეთის ფინანსური ზედმხედფელობის კომისია.
[9] Pensionfundsonline-Pension System In Poland-shorturl.at/ilA02

რაც შეეხება III დონეს, მოიცავს ნებაყოფლობით პროფესიულ პენსიებს. რაც გულისხმობს ე.წ. პროფესიული საპენსიო გეგმის შედგენას. თუმცა, განსხვავებით სხვა ღია საპენსიო ფონდებისაგან აღნიშნული პროფესიული საპენსიო სქემები გამოირჩევიან ინვესტირების შედარებით უფრო ფართო უფლებით. დამსაქმებლისთვის აღნიშნული სქემის შექმნა ნებაყოფლობითია და მისი ფინანსური კონტრიბუცია არ უნდა აღემატებოდეს დასაქმებულისთვის გადახდილი ხელფასის 7%-ს. გარდა ამისა, დასაქმებულსაც აქვს უფლება შეავსოს დამსაქმებლის მიერ გადებული შენატანი, თუმცა თავის მხრივ დასაქმებულიც შეზღუდული და ლიმიტირებულია გარკვეული ზღვრებით.

## 2. საპენსიო სისტემის გამოწვევები და განვითარების პერსპექტივები

### 2.1. დაგროვებითი საპენსიო პოლიტიკის არსი და არსებული მდგომარეობის შეფასება

მსოფლიო მასშტაბით ძალიან ბევრი ქვეყანა დადგა საპენსიო სისტემაში მნიშვნელოვანი და ზოგჯერ რადიკალური რეფორმების გატარების აუცილებლობის წინაშე. მსოფლიო ბანკისა და სავალუტო ფონდის ძირითადი მოთხოვნები ხშირ შემთხვევაში ეხება საპენსიო და სოციალური დაცვის სისტემების ხარჯების შემცირებას. სწორედ ამიტომ, ისეთმა განვითარებადმა ქვეყნებმაც კი, როგორიცაა ურუგუაი, კოლუმბია, ხორვატია და ა.შ. ძირეული ცვლილებები გატარეს საპენსიო სისტემებში.

თუმცა, გასათვალისწინებელია, რომ მსგავსი ცვლილებები ხშირ შემთხვევებში იწვევს მტკივნეულ ეკონომიკურ თუ პოლიტიკურ ცვლილებებსაც. თანამედროვე პირობებში საპენსიო სისტემებმა უნდა უპასუხონ არსებულ გამოწვევებს, რომელიც შეძლებს როგორც საბიუჯეტო რესურსებზე არსებული წნეხის შემცირებას და ასევე უზრუნველყოფს ადამიანების ღირსეულ სიბერეს.

სხვადასხვა წინაპირობების მიუხედავად ქვეყნები იძულებულნი ხდებიან დაგეგმონ და განახორციელონ საპენსიო სისტემები ყველა არსებული რისკის გათვალისწინებით, აღნიშნული კი გამოწვეულია მსოფლიო მოსახლეობის მატებით, შრომითი ძალის შემცირებით, მიმდინარე ეკონომიკური, ფინანსური, კრიზისებით. საპენსიო სისტემების სრულყოფის, რეფორმების გატარების მთავარი მიზანია შეიქმნას თანამედროვე, ლიბერალური და მდგრადი ეკონომიკური გარემო ქვეყანაში, რომელზეც შედარებით ნაკლები გავლენა ექნება საერთაშორისო დონეზე არსებულ ეკონომიკურ ცვლილებებს და მეტად გახშირებულ ფინანსურ კრიზისებს. სავალდებულო დაგროვებითი სისტემის დანერგვა კი თავის მხრივ, მნიშვნელოვან როლს შეასრულებს ქვეყნის ადეკვატური სოციალურ-ეკონომიკური პოლიტიკის ფორმირებაში აღნიშნული მიზნების გათვალისწინებით.

სავალდებულო დაგროვებითი საპენსიო სისტემის დანერგვამდე ქვეყანაში მოქმედებდა ორი საპენსიო სისტემა:

1. გადანაწილებით პრინციპზე დამყარებული საპენსიო სისტემა, რომლის მიხედვითაც პენსიების გაცემა ხდებოდა სახელმწიფო ბიუჯეტიდან.
2. კერძო საპენსიო უზრუნველყოფის სისტემები (დაინერგა 2001 წლიდან, კერძო საპენსიო დაზღვევა პირველმა ჯიპიაი ჰოლდინგმა შესთავაზა მოსახელობას), რომელიც საერთო ბაზრის ძალზედ უმნიშვნელო ნაწილს წარმოადგენდა. მიუხედავად არსებობის გარკვეული პერიოდისა მას არ გააჩნდა საკანონმდებლო ბაზა და განვითარების საწყის საფეხურებზე იმყოფებოდა.

ყველაზე მსხვილის კომპანია, რომელსაც გააჩნდა კერძო საპენსიო სისტემის ყველაზე მეტი ბენეფიციარი იყო ჯიპიაი ჰოლდინგი. საპენსიო ფონდის მოცულობა კი ყოველწლიურად იზრდებოდა:

**ცხრილი #4** ჯიპიაი ჰოლდინგის კერძო საპენსიო სისტემის ბენეფიციართა მოცულობის ზრდა წლების მიხედვით

| 2013 | 3 680 649,04 ₾ |
|------|----------------|
| 2014 | 4 108 071,28 ₾ |
| 2015 | 5 242 139,35 ₾ |
| 2016 | 5 906 471,09 ₾ |
| წყარო: ჯიპიაი ჰოლდინგი-ანგარიში | |

წყარო: ჯიპიაი ჰოლდინგი, 2016 წლის ინდივიდუალური ანგარიშგება[10]

აღნიშნული მატება ძალიან უმნიშვნელო, შესაბამისად შეგვიძლია თამამად ვთქვათ, რომ საქართველოში კერძო საპენსიო უზურნველყოფის სისტემებმა არ გაამართლა და პოპულარობა ვერ მოიპოვა საზოგადოებაში.

საქართველოში საპენსიო სისტემა ეფუძნება სოლიდარობის პრინციპს. აღნიშნული არ ითვალისწინებს პირის არც სამუშაო სტაჟს, არც შრომითი ურთიერთობების დროს მიღებულ შემოსავალს. თითოეული პენსიონერისთვის გაცემული ასაკობრივი პენსია თანაბარია. ცხადია, მსგავსი სისტემა ვერ იქნება მდგრადი. 2019 წლიდან ქვეყანაში დაინერგა ახალი, სავალდებულო დაგროვებითი საპენსიო სისტემა, რომელიც ძალიან ბევრ უცხოურ ქვეყანაში არსებობს.

მსოფლიოში დაგროვებითი საპენსიო სისტემის სხვადასხვა სახე არსებობს. მათგან ყველაზე წარმატებულია ჩილეს მაგალითი. ქვეყანაში 30 წელზე მეტია არსებობს დაგროვებითი პენსიის მოდელი, რომლის მიხედვითაც მოქალქეები სავალდებულო გადასახადს იხდიან ერთ-ერთი კერძო საპენსიო ფონდში. მათ მიერ შეტანილ თანხას კი ერიცხება გარკვეული საპროცენტო სარგებელი. ამასთან, არსებობს სხვადასხვა კერძო ფონდები, მოქალქეს შეუძლია მისი რესურსი მიმართოს ნებისმიერში, არსებული კონკურენცია კი იწვევს მაღალი საპროცენტო სარგებლის არსებობას.

თუმცა, გასათვალისწინებელია, რომ ჩილეს მაგალითი გამოსადეგია ქვეყნებში სადაც მოსახლეობას შედარებით მაღალი ანაზღაურება აქვთ. ბევრი ქვეყანა, რომელმაც მოდიფიცირების გარეშე განახორციელა ჩილეს მაგალითი ძალიან სწრაფადვე დაუბრუნდა სოციალურ პენსიებსა და განაწილებით სისტემას.

---

[10] https://www.gpih.ge/about/GEO2016_auditi.pdf

საქართველოში საპენსიო სისტემების რეფორმაზე მუშელობა დიდი ხნის წინ დაიწყო. 2018 წლიდან სახელმწიფო შეჯერდა ახალ საპენსიო სისტემაზე, რომელიც თავის მხრივ ითვალისწინებს მანამდე არსებულ პენსიას და პარალელურად სავალდებულო ხდება დაგროვებით საპენსიო სისტემაში[11] მონაწილეობა.

დაგროვებითი საპენსიო სისტემის მონაწილე მხარეები და მათ მიერ წარმოდგენილი შენატანები მოცემულია ცხრილში:

**ცხრილი #5** დაგროვებითი საპენსიო სისტემაში მონაწილეების პროცენტული შენატანები

| დამსაქმებელი | 2% |
|---|---|
| დასაქმებული, თვითდასაქმებული | 2% |
| სახელმწიფო | 2%, 1%, 0% |

წყარო: საქართველოს კანონი დაგროვებითი პენსიის შესახებ[12]

წლის განმავლობაში დარიცხული ხელფასის რაოდენობა თუ გადააჭარბებს 24 000 ლარს სახელმწიფოს კონტრიბუცია დამატებით თანხის 1%-თ განისაზღვრება, ხოლო თუ გადააჭარბებს 60 000 ლარს სახელმწიფო ფინანსური კონტრიბუციისგან თავისუფლდება.

დაგროვებითი საპენსიო სისტემა ამოქმედდა 2019 წლის 1 იანვრიდან გაცემულ ხელფასებზე. სისტემაში ჩართვის უფლება აქვს საქართველოს ყველა მოქალაქეს, ასევე საქართველოში მუდმივად მცხოვრებ სხვა ქვეყნის მოქალაქესაც. როგორც დასაქმებულებს, ასევე თვითდასაქმებულებსაც.

სისტემაში ჩართვა სავალდებულოა:

- 40 წლამდე ასაკის პირებისათვის
- დასაქმებულებისათვის, რომელთაც 2018 წლის 6 აგვისტოს 40 წელი არ ჰქონდათ შესრულებული.

მათ ვისაც 2018 წლის 6 აგვისომდე შეუსრულდათ 40 წელი სისტემაში ჩართვა ნებაყოფლობითია. დასაწყისისთვის ისინი ჩაერთვებიან სისტემაში, თუმცა 2019 წლის 1 აპრილიდან 31 მაისის ჩათვლით შეეძლოთ სისტემის დატოვება. ასევე ნებაყოფლობითია სისტემაში ჩართვა თვითდასაქმებული პირებისათვისაც. ამ შემთხვევაში ისინი საპენსიო ფონდში შემოსავლის 4%-ს შეიტანენ, ხოლო 2%-ს სახელმწიფო.

საპენსიო ასაკს მიდწევამდე შენატანების გატანა შესაძლებელია, თუ

- პირი შეიცვლის მოქალაქეობას და აღარ იქნება საქართველოს მოქალაქე
- გახდება შეზღუდული შესაძლებლობის მქონდე პირი
- გარდაცვალების შემთხვევაში საპენსიო შენატანების მფლობელი გახდება მემკვიდრე.

---
[11] 2018 წლის 21 ივლისის კანონი დაგროვებით პენსიის შესახებ.
[12] https://matsne.gov.ge/ka/document/view/4280127?publication=0

სავალდებულო დაგროვებითი საპენსიო სისტემის შექმნის აუცილებლობა და უპირატესობები:

სავალდებულო დაგროვებითი პენსიის ამოქმედებამდე მოქმედი საპენსიო სისტემის მთავარი ნაკლი იყო ის, რომ ხშირ შემთხვევაში მისი მოცულობა ვერ აკმაყოფილებდა პენსიონერის პირველად მოთხოვნილებებსაც კი. ანუ ის ვერ ასრულებდა მთავარ მისიას ხელი შეეწყო ხანდაზმულების ღირსეული სიბერისათვის. საქართველოში პენსიაზე გასვლის შემდგომ, საშუალოდ 5-ჯერ მცირდება ადამიანების შემოსავალი. ანუ ჩანაცვლების კოეფიციენტი ძალზედ დაბალია. დაგროვებითი საპენსიო სისტემის შექმნის მთავარი მიზანი, სწორედაც რომ ამ უსამართლობის აღმოფხვრაა.

აღნიშნული რეფორმის გატარების აუცილებლობას კიდევ უფრო ხაზს უსვამს სამუშაო ასაკში მყოფ ერთ პირზე პენსიონერთა რიცხვი. აღნიშნული კოეფიციენტი მიმდინარე პერიოდში 42%-მდე გაიზარდა. ნავარაუდებია, რომ ეს მონაცემი გარკვეული პერიოდის შემდგომ მნიშვნელოვნად გაიზრდება, შესაბამისად გაიზრდება საპენსიო ტვირთი დასაქმებულებზეც.

ყოველწლიურად იზრდება ბიუჯეტიდან საპენსიო უზრუნველყოფაზე მიმართული რესურსები. 2020 წლიდან კიდევ უფრო გაიზარდა ერთ პირზე გაცემული პენსიის მოცულობა. მოცემულ ცხრილში წარმოდგენილია მოსახლეობი საპენსიო უზრუნველყოფაზე გაწეული ხარჯები 2018-2020 წლებში:

ცხრილი #6 საპენსიო უზრუნველყოფაზე გაწეული საბიუჯეტო სახსრები 2018-2020 წლებში

| 2018 | 1,7 მლრდ |
|------|----------|
| 2019 | 1,9 მლრდ |
| 2020 | 2,3 მლრდ |
| წყარო: საქართველოს ბიუჯეტი 2020 | |

წყარო: საქართველოს ბიუჯეტი 2020[13]

2020 წელს საპენსიო უზრუნველყოფაზე გაწეული ხარჯი მთლიანი საბიუჯეტო ხარჯების 18,5%-ია. მოსახლეობის დაბერებისა და შობადობის კლების ფონზე სოლიდარობის პრინციპზე დამყარებული საპენსიო სისტემა ვერ უპასუხებს არსებულ გამოწვევებს. რეფორმების გარეშე წარმოუდგენელია თუნდაც არსებული საპენსიო განაკვეთის შენარჩუნება წლების განმავლობაში.

დაგროვებითი საპენსიო სისტემის მსგავსი მოდელი ძალიან ბევრ ქვეყანაშია გამოყენებული და დამკვიდრებული, მას არ გააჩნია ალტერნატივა ქვეყანაში ე.წ. გრძელი ფულის შექმნის თვალსაზრისით.გრძელვადიან პერიოდში მიღებული შედეგები ნამდვილად დიდი შეღავათი იქნება მოსახლეობისათვის, თუმცა უნდა აღვნიშნოთ ისიც, რომ მხოლოდ ამ სისტემის ამოქმედება არ იქნება საკმარისი ადამიანებისთვის ჯანსაღი ეკონომიკური თუ სოციალური გარემოს სრულყოფისათვის.

---

[13] https://mof.ge/5261

კულტურული ფაქტორი განაპირობებს იმას, რომ საქართველოს მოსახლეობის უმეტესობას ურჩევნია დახარჯოს ფული მიმდინარე პერიოდში ვიდრე გააკეთოს დანაზოგი. აღნიშნული განპირობებულია ფინანსური წიგნიერების დონის დაბალი მაჩვენებლითაც. ეკონომიკური განვითარებისა და თანამშრომლობის ქსელის (OECD) მიერ ჩატარებული კვლევის შედეგებს თუ დავეყრდნობით საქართველოს ფინანსური წიგნიერების მაქსიმალური 21 ქულიდან მხოლოდ 12,1 ქულა მიენიჭა. აღნიშნული კვლევა ნათლად ცხადყოფს საქართველოში ფინანსური განათლების დანერგვის აუცილებლობაზე. (იხ. ცხრილი #7)

ცხრილი #7: ქვეყნები წარმოდგენილია ფინანსური წიგნიერების დონის მიხედვით, OECD-ის კვლევა

| | Number of participants | Financial Literacy Score | Knowledge | Behaviour | Attitude |
|---|---|---|---|---|---|
| Austria | 1418 | 14.4 | 5.3 | 6.0 | 3.1 |
| Bulgaria | 1047 | 12.3 | 4.1 | 5.3 | 2.9 |
| Colombia | 1200 | 11.2 | 3.8 | 4.8 | 2.6 |
| Croatia | 1079 | 12.3 | 4.5 | 5.0 | 2.8 |
| Czech Republic | 1003 | 13.0 | 4.5 | 5.3 | 3.1 |
| Estonia | 1005 | 13.3 | 4.9 | 5.3 | 3.1 |
| Georgia | 1056 | 12.1 | 4.5 | 5.1 | 2.5 |
| Germany | 1003 | 13.9 | 5.2 | 5.7 | 3.1 |
| Hong Kong, China | 1002 | 14.8 | 6.2 | 5.8 | 2.9 |
| Hungary | 1001 | 12.3 | 4.6 | 4.5 | 3.3 |
| Indonesia | 1000 | 13.3 | 3.7 | 6.3 | 3.3 |
| Italy | 2036 | 11.1 | 3.9 | 4.2 | 3.0 |
| Korea | 2400 | 13.0 | 4.6 | 5.4 | 3.1 |
| Malaysia | 2818 | 12.5 | 3.7 | 6.1 | 2.7 |
| Moldova | 1074 | 12.6 | 4.0 | 5.5 | 3.1 |
| Montenegro | 1030 | 11.5 | 4.1 | 4.7 | 2.6 |
| Peru | 1205 | 12.1 | 4.1 | 5.1 | 2.9 |
| Poland | 1000 | 13.1 | 5.0 | 5.5 | 2.6 |
| Portugal | 1480 | 13.1 | 4.0 | 5.9 | 3.2 |
| North Macedonia | 1076 | 11.8 | 3.9 | 5.1 | 2.8 |
| Romania | 1060 | 11.2 | 3.5 | 5.0 | 2.7 |
| Russia | 83478 | 12.5 | 4.8 | 4.9 | 2.8 |
| Slovenia | 1019 | 14.7 | 4.8 | 6.3 | 3.6 |
| France * | 2155 | | 4.8 | | |
| Malta ** | 1013 | 10.3 | 2.2 | 5.2 | 2.8 |
| Thailand *** | 11129 | | 3.9 | | 3.9 |
| Average ^ | | 12.7 | 4.4 | 5.3 | 3.0 |
| Average (OECD-12) ^^ | | 13.0 | 4.6 | 5.3 | 3.1 |

წყარო: OECD, 2020 International Survey of Financial Literacy

საერთაშორისო პრაქტიკებზე დაყრდნობით, დაგროვებითი საპენსიო სისტემის სავალდებულო ხასიათი გრძელვადიანი პერიოდში გამოიწვევს დაგროვების კულტურის და დანაზოგის შექმნის უნარის გამომუშავებას.

კიდევ ერთი უპირატესობა რაც ახლავს ახალ სისტემას მისი ლიბერალური ხასიათია. საგადასახადო ტვირთი სრულიად დასაქმებულს არ აწვება და მას თანაბრად ინაწილებენ დამსაქმებლებიც და სახელმწიფოც. რამდენადაც ქვეყანაში დასაქმებული ადამიანების ხელფასი ხშირ შემთხვევაში მათ მოთხოვნილებებსაც კი ბოლომდე ვერ აკმაყოფილებს შემოთავაზებული სისტემა ადეკვატური და შესაბამისია.

სავალდებულო დაგროვებითი საპენსიო სისტემის შედეგი იქნება კაპიტალის ბაზრის ჩამოყალიბება. საპენსიო ფონდის მიერ მიმდინარე ეტაპზე მობილიზებული რესურსი შემდეგნაირად გამოიყურება:

ცხრილი #8: 2021 წლის მდგომარეობით საპენსიო ფონდში აკუმულირებული ფინანსური რესურსები

| საპენსიო ფონდში აკუმულირებული საპენსიო აქტივების ოდენობა (ლარი) | |
|---|---|
| 01.01.2020-ს მდგომარეობით | 508,030,682.48 |
| 01.01.2021-ს მდგომარეობით | 1,164,656,968.40 |
| 31.05.2021-ს მდგომარეობით | 1,483,567,052.33 |

**წყარო:** საპენსიო სააგენტო/გამოთხოვილი საჯარო ინფორმაცია ნავარაუდებია, რომ რეფორმიდან 6 წელიწადში აკუმულირებული რესურსის მოცულობა გადააჭარბებს 5 მლრდ ლარს. ეს ყოველივე ხელს შეუწყობს ქვეყანაში მანამდე არ არსებული კაპიტალის ბაზრის ჩამოყალიბებას და განვითარებას. ამ გზით კი ქვეყნის ეკონომიკური ზრდა შეადგენს 3-4%-ს.

დაგროვებითი საპენსიო სისტემის წარმატება ძალიან ბევრი ქვეყნის მაგალითზე შეგვიძლია განვიხილოთ, თუმცა საინტერესოა არის თუ არა მისი ეფექტი ერთი და იგივე ყველა სახელმწიფოში და გამართლებს თუ არა საქართველოში. ამასთან დაკავშირებით დაისვა არაერთი საფუძვლიანი შეკითხვა. მოსახლეობასა და ექსპერტებს არაერთგვაროვანი დამოკიდებულება გააჩნიათ. სისტემის მთავარი მიზნის მიღწევა-უზრუნველყოს მოსახლეობის ღირსეული სიბერე-კითხვითი ნიშნის ქვეშ დგება.

გაძვირებული სამუშაო ძალა – როგორც ცნობილია, დამსაქმებელმა საპენსიო ფონდში უნდა გადაიხადოს საშემოსავლო გადასახადით დაუბეგრავი სახელფასო ფონდის 2%. ეს შეიძლება დამსაქმებლებისთვის მძიმე ტვირთი იყოს, შესაბამისად ისინი ეცდებიან თავიდან აირიდონ გადახდა დასაქმებულთა ხელფასის კორექტირების (შემცირების) ხარჯზე. ცხადია, ამ შემთხვევაში იზღუდება დასაქმებულების უფლებებიც და ინტერესებიც. უნდა აღინიშნოს ასევე დასაქმების ბაზარზე გავლენა. კომპანიებს გაეზრდებათ შიდა ხარჯები რაც გამოიწვევს დამატებითი მუშახელის მოზიდვის ინტერესს. ქართულ რეალობაში საკმაოდ მძიმე სურათი გვაქვს დასაქმების ბაზარზე, სავალდებულო დაგროვებითი საპენსიო სისტემა კი შესაძლოა ამ პრობლემის გაღრმავების კატალიზატორად მოგვევლინოს.

რეფორმას თან ახლავს სავალუტო რისკებიც. აკუმულირებულ რესურსზე დარიცხულ სარგებელზე გავლენა შეიძლება იქონიოს სავალუტო რისკებმა. განვითარებულ ქვეყნებში მზარდი ეკონომიკისა და ინფლაციის დაბალი დონის პირობებში კერძო დაგროვებითი საპენსიო სისტემებიც კი ამართლებს. საქართველოში რადიკალურად განსხვავებული მდგომარეობა გვაქვს, მაღალია ინფლაციური და სავალუტო რისკები. აკუმულირებული რესურსის მსყიდველუნარიანობა შესაძლებელია სრულიად შეიცვალოს წლების შემდგომ. საბოლოოდ მაინც ვერ ვაღწევთ დასახულ მიზანს-ღირსეულ სიბერეს.

საპენსიო ფონდების პრაქტიკა აჩვენებს რომ ისინი როგორც წესი, ახორციელებენ საპორტფელო ინვესტიციებს, რაც გულისხმობს აქტივების ფასიან ქაღალდებში ინვესტირებას და არამც და არამც პირდაპირ ინვესტირებას. მიმდინარე მონაცემებით საპენსიო ფონდის საინვესტიციო პორტფელი ასე გამოიყურება:

ცხრილი#9: საპენსიო ფონდის საინვესტიციო პორტფელი

| | |
|---|---|
| სადეპოზიტო სერთიფიკატები | 1,11% |
| სარგებლიანი ანგარიშები | 30.19% |
| ვადიანი დეპოზიტები | 68.7% |

წყარო: საპენსიო სააგენტო[14]

საპენსიო სისტემის კრიტიკის მთავარი კრიტერიუმი მისი იძულებითი ბუნებაა. ადამიანს არ აქვს არჩევანი საკუთარი სახსრები გამოიყენოს, დახარჯოს თუ შეინახოს მომავლისთვის.

არაერთ ქვეყანაში საპენსიო სისტემების ჩამოშლის ფაქტი არსებობს. მათმა საქმიანობამ კი ფინანსური სკანდალიც კი გამოიწვია. ეს ყველაფერი ნდობის ფაქტორის არ არსებობაზე გამოიწვია საპენსიო ფონდესა და მოსახლეობას შორის. საქართველოს ძალიან ცუდი გამოცდილება გააჩნია, ე.წ. დაკარგული ანაბრების შემდგომ რთულია მოსახლეობის ნდობის მოპოვება.

შემდეგი ნაკლოვანება ეხება საპენსიო შენატანის ადმინისტრირებისა და დეკლარირების საკითხს. როგორც ვიცით, საგადასახადო კანონმდებლობა განაცემების შესახებ ინფორმაციის წარდგინებას მოითხოვს თვით დასრულებიდან არაუგვიანეს შემდეგი თვის 15 რიცხვისა. განსხვავებული მდგომარეობა გვაქვს საპენსიო შენატანებთან დაკავშირებით. ამ შემთხვევაში საპენსიო შენატანის დაკავების შესახებ სააგენტოს ინფორმაცია უნდა მიეწოდოს დარიცხვის განხორციელებისთანავე. განსაკუთრებით მცირე მეწარმეების შემთხვევაში აღრიცხვიანობა არ არის სისტემატიზირებული. აღნიშნული მოთხოვნები კი გადამხდელებს დამატებით ადმინისტრაციულ წნეხს უქმნის.

რეფორმის გატარებამდე მნიშვნელოვანია მოხდეს ყველა დეტალის შეფასება ანალიზი და სხვადასხვა კონსულტაციების გაევლა. საპენსიო რეფორმის დროს კომუნიკაციის არ არსებობამ სახელმწიფოსა და გადამხდელებს, და თვითონ სახელმწიფო უწყებებს შორის მნიშვნელოვან პრობლემებამდე მიგვიყვანა. ორაზროვანი იყო რიგი საკითხები, მათ შორის არაფულადი სარგებლის გაცემის შემთხვევაში როგორ დაითვლება საპენსიო შენატანი, ეხება თუ არა რეფრომა საპროცენტო შემოსავლებს და ა.შ.

### 2.2 დაგროვებითი საპენსიო სისტემის სრულყოფის მექანიზმები

სავალდებულო დაგროვებითმა საპენსიო სისტემამ მთელ რიგ ქვეყნებში გაამართლა და წარმატებით ფუნქციონირებს დღემდე. სისტემის ავკარგიანობის სრულიად განსასაზღვრად ჯერ ძალიან მცირე დრო გასულია, თუმცა არის საკითხთა ჩამონათვალი, რომელთა გათვალისწინების შემთხვევაში შესაძლოა მისი უკეთესი ვერსიით განვითარება.

---
[14] https://www.pensions.ge/investment-activity/investment-portfolio

წარმოდგენილ მასალაზე დაყრდნობით, განხილულია სისტემის სრულყოფისათვის რეკომენდაციები, რომელთა გათვალისწინების შემთხვევაში შესაძლებელი იქნება მოსალოდნელი რისკების თავიდან აცილება და სისტემის მოსარგებლეების კეთილდღეობის ზრდა.

პირველ რიგში მნიშვნელოვანია განისაზღვროს აკუმულირებული რესურსის განთავსების მიზნობრიობა და ეფექტიანობა. დღევანდელი მონაცემებით მისი 100% განთავსებულია ვადიან დეპოზიტებში და კომერციული ბანკების სარგებლიან ანგარიშებზე. შესაძლებელია გამოიძებნოს ბევრად უფრო მომგებიანი და უსაფრთხო ინვესტირების მექანიზმები.

სისტემაში ჩართულ პირებს უნდა მიეწოდოთ სრულყოფილი და სრულფასოვანი ინფორმაცია მათთვის გასაგებ და მარტივ ენაზე. თითოეული პროცესი უნდა იყოს გამჭირვალე, რამაც სისტემაში ნებაყოფლობით ჩართული პირების რაოდენობის ზრდაზეც შეიძლება იქონიოს გავლენა. საბოლოოდ კი ყოველივე აღნიშნული გაზრდის სისტემის სანდოობას.

მოსახლეობას მუდმივად და ინტენსიურად უნდა წარედგინოს ანგარიშები ინვესტირებული რესურსების მიმართულებების შესახებ. აუცილებელია საზოგადოებამ იცოდეს სად ინახება მათი რესურსი და რა სარგებელი ერიცხება მას. საინტერესო პრაქტიკაა ასევე, როცა საპენსიო სისტემი მონაწილე თავადვე ირჩევს შემოთავაზებული საინვესტიციო მიმართულებებიდან სად და როგორ განთავსდეს საკუთარი რესურსი. იზრდება პასუხისმგებლობის განცდა, რომელსაც საპენსიო სააგენტოსთან ერთად ინაწილებს სისტემის მოსარგებლე.

ინვესტირების დროს ძალიან დიდი მნიშვნელობა ენიჭება ინფლაციის დონის გათვალისწინებას. ინფლაციის ინდექსაცია კარგი გამოსავალი შეიძლება იყოს.
აგრეთვე, უმნიშვნელოვანესია, რომ მოხდეს საპენსიო სააგენტოს მმართველი რგოლის საქმიანობის მუდმივი მონიტორინგი და რეგულირება რადგან თავიდან ავიცილოთ მათ მიერ არამართლზომიერი რისკების აღება და საპენსიო სისტემის ბენეფიციარების დაზარალება.

სისტემის შემოღების შემდეგ აუცილებელია თანმდევი რეფორმები განხორციელდეს და დაიხვეწოს სხვადასხვა მიმართულება როგორიცაა: ფასიანი ქაღალდების ბაზარი, საპენსიო შენატანების დაზღვევა და ა.შ. საჭიროა ეფექტურად განხორციელედეს რეფორმა, რათა კაპიტალის ბაზრის განვითარებას შეუწყოს ხელი. საბოლოოდ კი ქვეყნის მაკროეკონომიკური მაჩვენებლები სასურველ ნიშნულებს მიუახლოვდეს.

რამდენადაც საქართველოს ბაზარზე შეზღუდულია ფინანსური ინსტრუმენტები, საწყის ეტაპზეა კორპორატიული მართვაც, გვაქვს კვალიფიციური კადრების ნაკლებობა და ამასთან, მცირე ბაზრის ზომაც, უფრო დიდ მნიშვნელობას იძენს საპენსიო რეფორმის წარმატებით დანერგვა.

საქართველოს არ უნდა გაიმეოროს პოლონეთის შემთხვევა, სადაც სავალდებულო საპენსიო სისტემის დანერგვის მიუხედავად მაინც ვერ განვითარდა კაპიტალის ბაზარი და საწყის ეტაპებს ვერ გასცდა და ძალიან მცირე რაოდენობის ვაჭრობა ფიქსირდება. საპენსიო სააგენტოს უნდა ჰქონდეს მუდმივი კავშირი სისტემის მონაწილე არა მარტო ფიზიკურ, არამედ იურიდიულ პირებთანაც.

აუცილებლად გასათვალისწინებელია პოლიტიკური რისკების მნიშვნელობაზე ხაზგასმაც. სხვადასხვა ფისკალური თუ ფინანსური პრობლემების პერიოდში მსოფლიო პრაქტიკა გვიჩვენებს, რომ შესაძლოა საპენსიო ფონდში აკუმულირებული რესურსები სხვადასხვა მიზნობრიობით გაიხარჯოს. ამის ძალიან კარგი მაგალითები არიან არგენტინა, უნგრეთი სადაც 2008-2010 წლებში საბიუჯეტო რესურსების ნაკლებობის გამო მოახდინეს კერძო საპენსიო ფონდების ნაციონალიზაცია. სომხეთში სავალდებულო დაგროვებითი საპენსიო სისტემა 2014 წელს დაინერგა, თუმცა მწვავე დებატების და კამათის შემდგომ სასამართლომ არაკონსტიტუციურად სცნო რეფორმა და მისი განხორციელება შეჩერდა. ყოველივე ზემოთ ჩამოთვლილისგან თავის ასარიდებლად სახელმწიფომ უნდა უზრუნველყოს მოსახლეობისთვის გარანტის მიცემა.

რამდენადაც რეფორმა შეიძლება ითქვას რომ ჯერჯერობით ისევ საწყის ეტაპზეა, საპენსიო სააგენტო უნდა ცდილობდეს მუდმივ და აქტიურ რეჟიმში მოახდინოს პრობლემათა იდენტიფიცირება და შეძლოს რისკების მინიმუმამდე დაყვანა.
და ბოლოს, ძალიან მკაფიოდ უნდა იქნას განსაზღვრული საპენსიო აქტივების ინვესტირების მექანიზმები.

დასკვნა
1. ამრიგად, საპენსიო პოლიტიკა ნებისმიერი ქვეყნის სოციალურ-ეკონომიკური მდგარადობის ერთ-ერთი მთავარი განმაპირობებელი ფაქტორია. ხანდაზმულთა რაოდენობის ზრდამ და თანამედროვე დემოგრაფიულმა პრობლემებმა მნიშვნელოვანი კითხვითი ნიშნები გააჩინა გადანაწილებითი საპენსიო სისტემის მიმართულებით.საქართველოში დღევანდელი საპენსიო სისტემის ჩამოყალიბება მჭიდრო კავშირშია ეტაპობრივად განხორციელებულ ამა თუ იმ რეფორმასთან, რომლებიც წლების განმავლობაში ინერგებოდა.
2. საპენსიო სისტემების სრულყოფის, რეფორმების გატარების მთავარი მიზანია შეიქმნას თანამედროვე, ლიბერალური და მდგრადი ეკონომიკური გარემო ქვეყანაში, რომელზეც შედარებით ნაკლები გავლენა ექნება საერთაშორისო დონეზე არსებულ ეკონომიკურ ცვლილებებს და მეტად გახშირებულ ფინანსურ კრიზისებს.საქართველოში
3. 2019 წლიდან გატარდა მნიშვნელოვანი რეფორმა სავალდებულო დაგროვებით საპენსიო სისტემასთან დაკავშირებით. აღნიშნულმა ცვლილებამ გამოიწვია რიგი განხილვები და დებატები ქვეყანაში.იმისთვის, რომ სისტემის დადებით შედეგებზე ვისაუბროთ საჭიროა რეფორმიდან რამდენიმე ათეული წელი მაინც გავიდეს.
4. ქვეყანაში ძალიან მძიმე სოციალური ფონია შექმნილი, მოსახლეობა მუდმივად ცდილობს სიღარიბის დაძლევის ეფექტური გზების შერჩიოს. მდგრადი და წარმატებული საპენსიო სისტემის შექმნა შეიძლება ითქვას ერთგვარი პოლიტიკური იარაღია ამ ბრძოლაში წარმატების მისაღწევად.

**References**

1. გ. სეფიაშვილი, საქართველოს საპენსიო სისტემა-არსებული გამოწვევები და პერსპექტივები.
2. პენსიის მიმღებთა საერთო რაოდენობა, საქართველოს სოციალური მომსახურების სააგენტო, shorturl.at/cxIT0
3. საპენსიო სააგენტო, წლიური ანგარიში, shorturl.at/fwxBD
4. საკანონმდებლო მაცნე, საქართველოს კანონი დაგროვებითი პენსიის შესახებ, shorturl.at/czBN7
5. თ. აბსანიძე, საპენსიო რეფორმა-ღირსეული სიბერის გარანტი თუ ახალი საგადასახადო ტვირთი?!, shorturl.at/tPSZ8.
6. ტაბულა, ახალი საპენსიო რეფორმა, http://tbl.ge/2mhj
7. საქართველოს ეკონომიკისა და მდგრადი განვითარების სამინისტროს ანგარიში, საპენსიო რეფორმა, shorturl.at/kFR79.
8. განმარტებითი ბარათი საქართველოს კანონ პროექტზე „დაგროვებითი პენსიის შესახებ"
9. ქ. ესერიძე, საპენსიო პოლიტიკის ევოლუცია საქართველოში, 2019
10. საერთაშორისო გამჭირვალობა, შენიშვნები საპენსიო რეფორმაზე, shorturl.at/ivxCT.
11. ნ. ხელაია, საპენსიო პოლიტიკის ევოლუცია: რატომ გახდა საჭირო ევროპის ქვეყნებში საპენსიო სისტემების რეფორმირება, 2018, shorturl.at/mpPTV.
12. შ. ტყეშელაშვილი, ფინანსური განათლების დონე საქართველოში-OECD-ის კვლევა, 2020, shorturl.at/uRW59.
13. მ. ჰუთსებაუთი, საპენსიო სისტემის რეფორმა საქართველოში, შენიშვნები და რეკომენდაციები, 2017.
14. Abuselidze, G. (2018). Optimal Fiscal Policy–Factors for the Formation of the Optimal Economic and Social Models. J. Bus. Econ. Review, 3(1), 18-27.
15. Abuselidze, G., & Slobodyanik, A. (2018). Overview welfare in Georgia and prospects of formation of effective economic and social models. Сучасні питання економіки і права [Modern Questions of Economics and Law], 2(8).
16. OECD/INFE 2020 International Survey of Adult Financial Literacy, shorturl.at/wNXY3.
17. A World Bank Policy Research Report, Averting the Old Age Crisis, shorturl.at/Rdelp.
18. Pensions Systems by Country, Country Profiles, https://www.pensionfundsonline.co.uk/content/country-profiles/.